\documentclass[10pt,twoside,a4paper]{amsart}
\usepackage{amsmath}
\usepackage{amsfonts}
\usepackage{amssymb}
\usepackage{amsthm}
\usepackage{newlfont}
\usepackage{graphicx}
\usepackage{amscd}

\theoremstyle{plain}
\newtheorem{Th}{Theorem}[section]
\newtheorem{Cor}[Th]{Corollary}
\newtheorem{Lem}[Th]{Lemma}
\newtheorem{Prop}[Th]{Proposition}
\theoremstyle{definition}
\newtheorem{Def}{Definition}[section]

\theoremstyle{remark}
\newtheorem*{Rem}{Remark}
\numberwithin{equation}{section}
\newcommand{\PP}{{\mathbb P}}
\newcommand{\EE}{{\mathbb E}}
\newcommand{\CC}{{\mathbb C}}
\newcommand{\FF}{{\mathbb F}}
\newcommand{\ZZ}{{\mathbb Z}}
\newcommand{\RR}{{\mathbb R}}
\newcommand{\AAf}{{\mathbb A}}
\newcommand{\bx}{\boldsymbol{x}}
\newcommand{\by}{\boldsymbol{y}}

\newcommand{\bt}{\boldsymbol{t}}

\newcommand{\bY}{{\boldsymbol Y}}

\newcommand{\bX}{\boldsymbol{X}}
\newcommand{\tbX}{\tilde{\bX}}

\newcommand{\tv}{\tilde{v}}
\newcommand{\tH}{\tilde{H}}
\newcommand{\tQ}{\tilde{Q}}

\newcommand{\D}{{\Delta}}
\newcommand{\tD}{\tilde{\D}}
\newcommand{\tA}{\tilde{A}}

\begin{document}

\title[The C-quadrilateral lattice]
{The C-(symmetric) quadrilateral lattice, its 
transformations 
and the algebro-geometric construction}

\author{Adam Doliwa}

\address{A. Doliwa, Wydzia{\l} Matematyki i Informatyki,
Uniwersytet Warmi\'{n}sko-Mazurski w Olsztynie,
ul.~\.{Z}o{\l}nierska 14, 10-561 Olsztyn, Poland}

\email{doliwa@matman.uwm.edu.pl}

%

\begin{abstract}
The C-quadrilateral lattice (CQL), called also the symmetric lattice, 
provides geometric interpretation of the
discrete CKP equation within the quadrilateral lattice (QL) theory. We discuss
affine-geometric properties of the lattice emphasizing the role of the Gallucci
theorem in the multidimensional consistency of the CQL. Then we give the
algebro-geometric construction of the lattice. We also present the reduction of
the vectorial fundamental transformation of the QL to the CQL case. In the
Appendix we show a relation between the QL and the so called Darboux maps. 
\end{abstract} 
\maketitle

\tableofcontents

\section{Introduction}
The difference equations play an increasing role in science. We have
learned to appreciate the inherent discreteness of physical phenomena at the
atomic and subatomic level. Also in modern theoretical 
physics the assumption of
a space-time continuum is often being abandoned. Apart from physics 
difference equations have 
numerous applications, e.g., in numerical analysis, computer science, 
mathematical biology and economics.

\begin{figure}
\begin{center}
\includegraphics{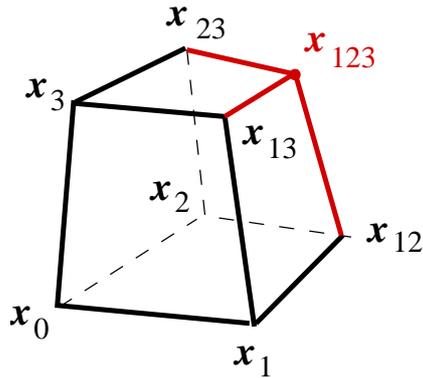}
\end{center}
\caption{The geometric integrability scheme}
\label{fig:TiTjTkx}
\end{figure}

The domain of discrete (difference) systems forms nowadays one of the focal
points in integrable systems research. Also the connection between geometry
and integrability, well known in the case of integrable differential equations
\cite{Sym,RogersSchief,GuHuZhou}, has been transferred to the discrete level 
(see \cite{DS-EMP} and references therein). 
A successful general approach towards description of this relation 
is provided by the theory of
multidimensional quadrilateral lattices (QLs)  \cite{MQL}. These are just
maps $x:\ZZ^N\to\PP^M$ ($3\leq N\leq M$) of multidimensional integer lattice
into projective space, with planar elementary quadrilaterals.
The integrable partial difference equation counterpart of the QLs are
the discrete Darboux equations (see Section \ref{sec:MQL-affine} for details), 
being found first \cite{BoKo} as the most general
difference system integrable by the non-local 
$\bar\partial$ dressing method. 
It turns out that integrability of the discrete Darboux system is encoded in a
very simple geometric statement (see Fig.~\ref{fig:TiTjTkx}). 
\begin{Lem}[The geometric integrability scheme] \label{lem:gen-hex}
Consider points $x_0$, $x_1$, $x_2$ and $x_3$ in general position in $\PP^M$,
$M\geq 3$. On
the plane $\langle x_0, x_i, x_j \rangle$, $1\leq i < j \leq 3$ choose a point
$x_{ij}$ not on the lines  $\langle x_0, x_i \rangle$, $\langle x_0,x_j
\rangle$ and $\langle x_i, x_j \rangle$. Then there exists the
unique point $x_{123}$
which belongs simultaneously to the three planes 
$\langle x_3, x_{13}, x_{23} \rangle$,
$\langle x_2, x_{12}, x_{23} \rangle$ and
$\langle x_1, x_{12}, x_{13} \rangle$.
\end{Lem}

Integrable reductions of the quadrilateral lattice (and thus of the discrete
Darboux equations) arise from additional constraints which are compatible with
geometric integrability scheme (see, for example \cite{q-red,DS-sym,BQL}). 
Because application of several integrable reductions
preserves integrability of the lattice, it is important to isolate the basic
ones. 
Integrable constraints of the quadrilateral lattice 
may have local or global nature. 
The global constraints are related with existence of an
additional geometric structure in the projective space which allows to impose
some restrictions on the lattice. 
The reduction considered in this paper can be described
within the affine geometry approach.

The differential
Darboux equations, which have appeared first in projective
differential geometry of multidimensional conjugate nets \cite{Darboux-OS}, 
play an important role \cite{BoKo-N-KP} in the multicomponent
Kadomtsev--Petviashvilii (KP) hierarchy, 
which is commonly considered \cite{DKJM,KvL}
as the fundamental system of equations in integrability theory.
One of the most important reductions of the KP hierarchy of nonlinear equations
is the so called CKP hierarchy \cite{DJKM-CKP} 
(here ``C" appears in the context of the classification theory of simple 
Lie algebras). In \cite{AvL} it was shown that the differential Darboux
equations of the multicomponent KP hierarchy should be 
in this case supplemented by certain symmetry  condition.
 In fact such equations were considered first 
within the differential geometry context in \cite{Darboux-OS,Bianchi}. Their
discrete counterpart was studied in \cite{DS-sym} under the name of the 
symmetric quadrilateral lattice. In \cite{DS-sym} the symmetric Darboux
equations have been solved by the non-local
$\bar\partial$ dressing method. The corresponding reduction of the fundamental
transformation \cite{TQL} of QL was given on the algebraic level
in \cite{MM}. 

Within the context of the so
called Darboux maps such a reduction of the discrete Darboux equations was
studied in \cite{Schief-JNMP}, where the system was reformulated in 
a convenient scalar form (see also \cite{DJM})
and related with the superposition formulas for the Darboux
transformations of the CKP hierarchy. In \cite{KingSchief} such special Darboux
maps were characterized geometrically using the classical Pascal's theorem for
hexagons inscribed in conics. Also the B\"{a}cklund transformation for such
maps was investigated geometrically in \cite{KingSchief}.
In fact,
the attempt to understand results of \cite{Schief-JNMP,KingSchief} in the
quadrilateral lattice approach was my
motivation to undertake again, after \cite{DS-sym}, 
studies of the symmetric reduction of the Darboux equations.

An important reduction of the quadrilateral lattice with the symmetric rotation
coefficients is the Egorov lattice \cite{Schief-Egorov}. It was solved in
\cite{DS-sym} using the inverse spectral transform. In \cite{AKV} the
algebro-geometric techniques were used to construct large class of such
lattices. In 1998 Peter Grinevich 
isolated from \cite{AKV} these of the algebro-geometric conditions 
which give rise the symmetric quadrilateral lattices \cite{PG-private}.

The main goal of the paper is to give new pure geometric characterization of 
the symmetric quadrilateral lattice and of the
corresponding reduction of the fundamental transformation. 
Because of the intimate relation of the lattice with
the discrete CKP equation we will call it also the C-quadrilateral lattice.
We study the geometric and algebraic properties of the CQL reduction of the
fundamental transformation. In particular, we show that the
reduced transformation satisfies the Bianchi permutability principle. To
make the theory of such lattices complete we apply the algebro-geometric
techniques to construct large classes of the lattices together with
corresponding solutions of the Darboux equations. In doing that we introduce
the
backward and the dual Baker--Akhiezer functions of the quadrilateral lattice. 

The paper is organized as follows. In Section \ref{sec:MQL-affine} we summarize
relevant facts from the quadrilateral lattice theory and we introduce the notation.
In Section \ref{sec:CQL} we introduce new geometric definition of the
C-quadrilateral lattice and we study its integrability (multidimensional
consistency) using geometric means. We
also present its algebraic description in terms of symmetric Darboux
equations. Section \ref{sec:AG} is devoted to presentation of the
algebro-geometric construction of the C-quadrilateral (symmetric) lattices. 
In Section \ref{sec:C-fund} we introduce geometrically the C-reduction
of the fundamental transformation of the quadrilateral lattice \cite{TQL} and we
link it with the earlier algebraic results of \cite{MM}. We also prove the
corresponding permutability theorem for this transformation. Finally in the
Appendix we present briefly
the relation between the Darboux maps \cite{Schief-JNMP} and the
quadrilateral lattices.

The main geometric results of the paper were presented on the 
LMS Research Symposium  \emph{Methods of Integrable Systems in Geometry}, 
Durham, UK (August, 2006). It is my pleasure to thank the organizers of the
Meeting for invitation and support.

\section{The multidimensional quadrilateral lattice (affine description)}
\label{sec:MQL-affine}
\subsection{The discrete Laplace and Darboux equations}
Consider a multidimensional quadrilateral lattice (MQL); i.~e., a mapping
$x :\ZZ^N \rightarrow \PP^M(\FF)$, $3\leq N\leq M$,
with all the elementary quadrilaterals planar \cite{MQL}; here $ZZ^N$ is the
$N$-dimensional integer lattice, and $\PP^M(\FF)$ is $M$-dimensional projective
space over the field $\FF$. 
In the affine gauge (which will be extremely useful in the paper) the 
lattice is represented by a mapping
$\bx: \ZZ^N \rightarrow \FF^M$ and the planarity
condition can be formulated in terms of the Laplace equations
\begin{equation}  \label{eq:Laplace}
\D_i\D_j\bx=(T_{i} A_{ij})\D_i\bx+
(T_j A_{ji})\D_j\bx, \qquad i\not= j, \qquad  i,j=1 ,\dots, N,
\end{equation}
where $T_i$ is the translation operator in the $i$-th direction, and 
$\D_i = T_i - 1$ is the corresponding partial difference operator. 

Due to compatibility of the system \eqref{eq:Laplace} 
the coefficients $A_{ij}$ satisfy the MQL (or discrete Darboux) system of
equations
\begin{equation} \label{eq:MQL-A}
\D_k A_{ij} + (T_kA_{ij})A_{ik} =
 (T_jA_{jk})A_{ij} +(T_k A_{kj})A_{ik} ,
\qquad i, j, k \quad \text{distinct}.
\end{equation}
The $j\leftrightarrow k$ symmetry of RHS of \eqref{eq:MQL-A} implies existence
of the potentials $H_{i}$, $i=1, \dots , N$, 
(called the Lam\'e coefficients) such that
\begin{equation}   \label{def:A-H}
A_{ij}= \frac{\D_j H_i}{H_i} \; , \qquad i\ne j \; .
\end{equation}
If we introduce the suitably scaled tangent 
vectors $\bX_i$, $i=1,...,N$,
\begin{equation}  \label{def:HX}
\D_i\bx = (T_iH_i) \bX_i,
\end{equation}
then equations \eqref{eq:Laplace} can be rewritten as a
first order system expressing the fact that $j$th variation of $\bX_i$ is
proportional to $\bX_j$ only
\begin{equation} \label{eq:lin-X}
\D_j\bX_i = (T_j Q_{ij})\bX_j,    \qquad i\ne j \; .
\end{equation}
The proportionality factors $Q_{ij}$, called the rotation coefficients, 
can be found from the linear equations
\begin{equation} \label{eq:lin-H}
\D_iH_j = (T_iH_i) Q_{ij}, \qquad i\ne j \; ,
\end{equation}
adjoint to \eqref{eq:lin-X}.
The compatibility condition for the system~\eqref{eq:lin-X} (or its
adjoint)
gives the following new form of the discrete Darboux equations
\begin{equation} \label{eq:MQL-Q}
\D_kQ_{ij} = (T_kQ_{ik})Q_{kj}, \qquad i\neq j\neq k\neq i.
\end{equation}

\subsection{The backward data, the connection factors and the $\tau$-function}
The backward tangent vectors $\tbX_i$, $i=1,\dots,N$, are defined similarly to the
forward tangent vectors $\bX_i$ but with the help of the backward
shifts $T_i^{-1}$ and of the backward
difference operator $\tD_i := 1- T_i^{-1}$:
\begin{equation} \label{eq:lin-bX}
\tD_i\tbX_j = (T_i^{-1} \tQ_{ij})\tbX_i \; , \qquad \text{or}
\qquad \D_i\tbX_j =  (T_i\tbX_i)\tQ_{ij},    \quad i\ne j ;
\end{equation}
we define also the backward rotation coefficients $\tQ_{ij}$ as the
corresponding proportionality factors. Notice that the backward tangent 
vectors $\tbX_i$ satisfy the adjoint linear system \eqref{eq:lin-H}.
\begin{figure}
\begin{center}
\includegraphics[width=8cm]{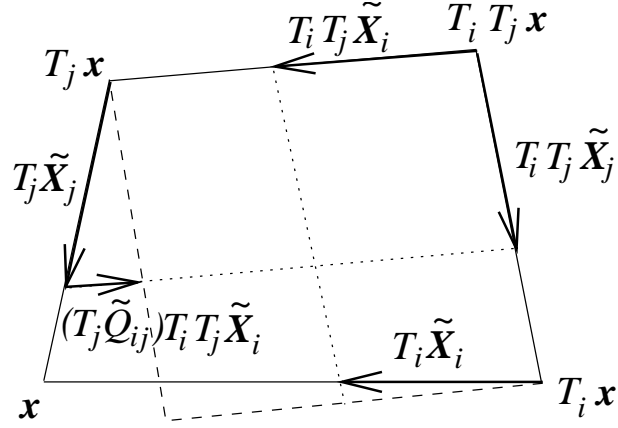}
\end{center}
\caption{Definition of the backward data}
\label{fig:back}
\end{figure}

The backward Lam\'e coefficients $\tH_i$, , $i=1,\dots,N$, are defined by
\begin{equation} \label{eq:b-H-X}
\tD_i\bx = (T_i^{-1}\tH_i ) \tbX_i \; , \qquad \text{or}
\qquad \D_i\bx = \tH_i (T_i\tbX_i ) \, ,
\end{equation}
and satisfy the system 
\begin{equation} \label{eq:lin-bH}
\tD_j\tH_i =  (T_j^{-1}\tH_{j})\tQ_{ij},  \qquad \text{or} 
\qquad \D_j\tH_i =  (T_j\tQ_{ij})\tH_{j},   \quad i\ne j \; .
\end{equation}
As a consequence of equations \eqref{eq:lin-bX} or equations \eqref{eq:lin-bH}
the functions $\tQ_{ij}$ satisfy the MQL 
equations~\eqref{eq:MQL-Q}.
Define backward Laplace coefficients
\begin{equation}
\tA_{ij} = \frac{\tD_j \tH_i}{\tH_i}\; ,
\end{equation}
then $\bx$ satisfies the backward Laplace equation
\begin{equation} \label{eq:Laplace-b}
\tD_i\tD_j\bx = (T_i^{-1}\tA_{ij})\tD_i\bx +
(T_j^{-1}\tA_{ji})\tD_j\bx
\; , \qquad i\ne j \; .
\end{equation}

The connection factors $\rho_i:\ZZ^N\to \FF$ are the 
proportionality coefficients between
$\bX_i$ and $T_i\tbX_i$ (both vectors are proportional to $\D_i\bx$):
\begin{equation} \label{eq:def-rho}
\bX_i = \rho_i ( T_i\tbX_i) \; , \qquad
i=1,\dots ,N \; .
\end{equation}
Other forward and backward data of the lattice $\bx$ are related 
through the following formulas
\begin{align} \label{eq:Q-Qt}
\tH_i & =  \rho_i  T_iH_i , \\
\rho_j T_j\tQ_{ij} & =  \rho_i T_iQ_{ji} \; .
\end{align}
Moreover, the factors $\rho_i$ are satisfy
equations
\begin{equation} \label{eq:rho-constr}
\frac{T_j\rho_i}{\rho_i} = 1 - (T_iQ_{ji})(T_jQ_{ij}), \qquad  i\ne j \; ,
\end{equation}
which imply existence of yet another potential
(the $\tau$-function of the quadrilateral lattice) such that
\begin{equation} \label{eq:tau}
\rho_i = \frac{T_i \tau}{\tau} .
\end{equation}

Given a quadrilateral lattice $\bx$, the forward data $\{\bX_i, H_i, Q_{ij}\}$
and the backward data $\{\tilde\bX_i, \tilde{H}_i, \tilde{Q}_{ij}\}$
are defined up to rescaling by functions $a_i(m_i)$, $b_i(m_i)$ of single
variables
\begin{gather} \label{eq:forward-scaling}
\bX_i \to a_i\bX_i, \qquad T_i H_i \to \frac{1}{a_i} T_i H_i , \qquad
T_j Q_{ij} \to \frac{a_i}{a_j} T_j Q_{ij}, \\
\label{eq:backward-scaling}
T_i\tilde\bX_i \to \frac{1}{b_i}T_i\tilde\bX_i, 
\qquad \tilde{H}_i \to b_i\tilde{H}_i, \qquad
T_j \tilde{Q}_{ij} \to \frac{b_i}{b_j} T_j \tilde{Q}_{ij},
\end{gather}
then also
\begin{equation}
\rho_i \to a_i b_i \rho_i.
\end{equation}

\begin{Rem}
Since $Q_{ij}$ and $\tQ_{ij}$ are both solutions of the discrete Darboux
equations \eqref{eq:MQL-Q}, then equations 
\eqref{eq:def-rho}-\eqref{eq:rho-constr} describe a special symmetry
transformations of \eqref{eq:MQL-Q}, first found in \cite{KoSchief2} without any
associated geometric meaning.
\end{Rem}

\section{The C-quadrilateral lattice}
\label{sec:CQL}
\subsection{Geometric definition of the C-quadrilateral lattice}
The geometric arena, where the reduction of the quadrilateral lattice studied in
this paper lives, is the affine space. Recall that
the affine transformations are these
projective transformations which leave invariant a fixed hyperplane 
$H_\infty\subset\PP^M$, called the hyperplane at infinity (see, for example
\cite{Coxeter-PG}). Two lines of $\AAf^M = \PP^M \setminus H_\infty$ 
(identified with $FF^M$, up to
fixing the origin) are 
called parallel if they intersect in a point of $H_\infty$

\begin{figure}
\begin{center}
\includegraphics[width=12cm]{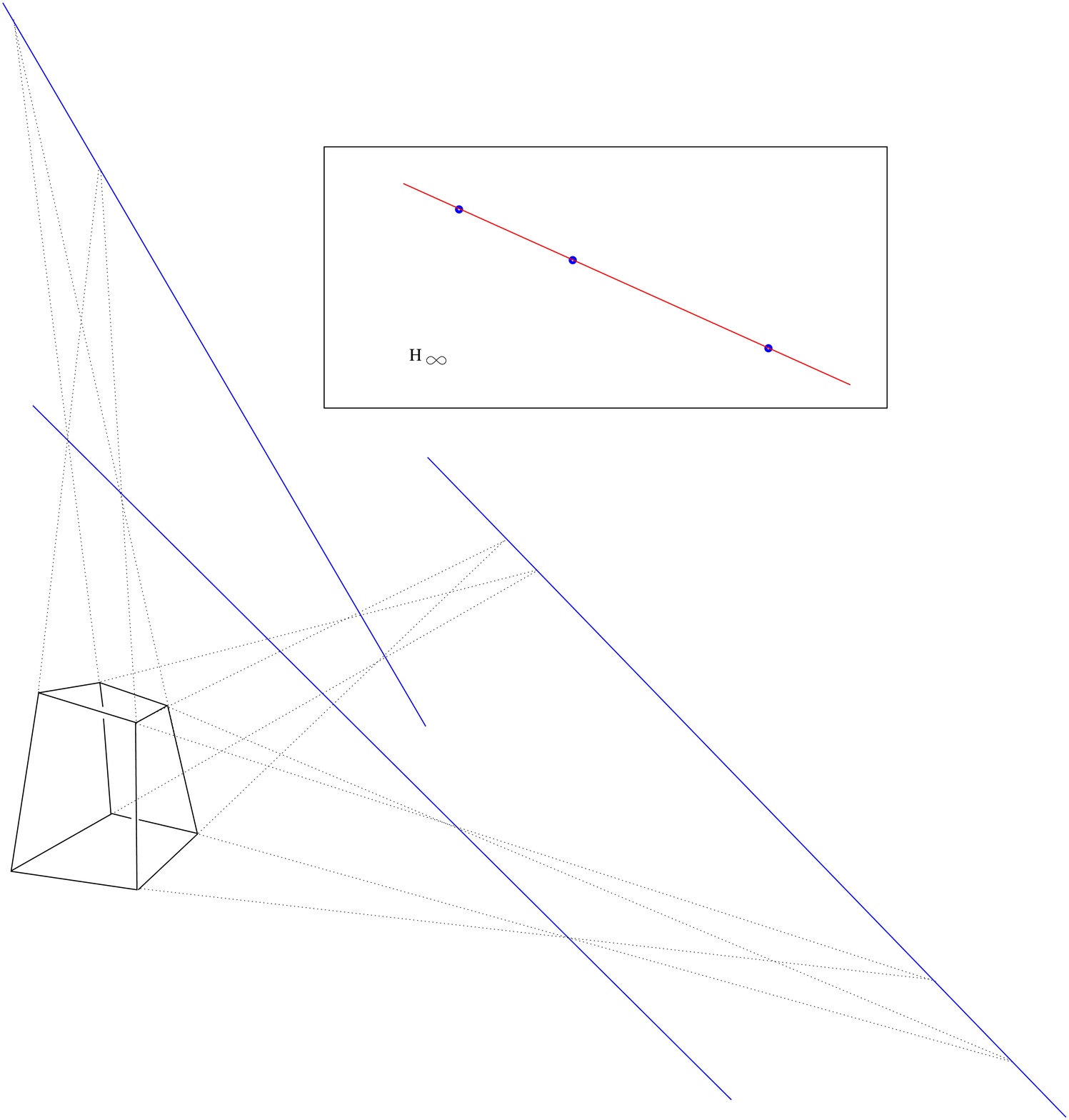}
\end{center}
\caption{The C-hexahedron}
\label{fig:CQL-constr}
\end{figure}

\begin{Def} \label{def:C-hexahedron}
A hexahedron with planar faces in the affine space
$\AAf^M$ is called a \emph{C-hexahedron} if the three points obtained by
intersection of the common lines of the pairs of planes of its opposite 
faces with the hyperplane $H_\infty$ at infinity 
are collinear (see Figure~\ref{fig:CQL-constr}).
\end{Def}
\begin{Rem}
We exclude for a time being
from our considerations the degenerate case when 
a pair of opposite faces of the hexahedron is parallel, i.~e. their common
line belongs to $H_\infty$.
\end{Rem}
\begin{Cor}
Equivalently  the C-hexahedron is characterized by coplanarity of
suitable parallel shifts of
the intersection lines of the opposite face planes.
\end{Cor}
\begin{Def} \label{def:C-hex}
A quadrilateral lattice $\bx:\ZZ^N\to\AAf^M$ is called a
\emph{C-quadrilateral lattice} (CQL) if all its hexahedra are C-hexahedra.
\end{Def} 
The following Proposition gives an analytic characterization of C-quadrilateral
lattices in terms of their rotation coefficients.
\begin{Prop} \label{prop:Q3-CQL}
A quadrilateral lattice is subject to C-reduction if and only if its 
rotation coefficients satisfy the constraint
\begin{equation} \label{eq:CQL-constr}
(T_jQ_{ij}) (T_kQ_{jk}) (T_iQ_{ki}) = (T_jQ_{kj}) (T_kQ_{ik}) (T_iQ_{ji}), 
\qquad i, j, k \quad \text{distinct}.
\end{equation}
\end{Prop}
\begin{proof}
Denote by $\bt^k_{ij}$ ($i,j,k$ are
distinct) the direction vector of the common line of the plane 
$\langle \bx, T_i \bx, T_j\bx \rangle$ and its $k$-opposite 
$\langle T_k\bx, T_i T_k \bx, T_j T_k \bx \rangle$. 
It must be therefore decomposed in
the basis $\{ \bX_i,\bX_j \}$ and in the basis $\{ T_k\bX_i,T_k \bX_j \}$.
Assuming its decomposition in the second basis we get
\begin{equation*}
\bt^k_{ij} = a T_k\bX_i + b T_k \bX_j = a \bX_i + b \bX_j +
(a T_k Q_{ik} + b T_k Q_{jk} ) \bX_k,
\end{equation*}
where we have used the linear problem \eqref{eq:lin-X}. Because the coefficient
in front of $\bX_k$ must vanish, the vector can be therefore
chosen as
\begin{equation}
\bt^k_{ij}  = T_k Q_{jk} \bX_i - T_k Q_{ik} \bX_j . 
\end{equation}
Notice that the condition in Definition~\ref{def:C-hex} is equivalent to  
to coplanarity of the vectors  $\bt^k_{ij}$, $\bt^j_{ik}$ and 
$\bt^i_{jk}$, and the statement follows from equation
\begin{equation*}
\bt^k_{ij}\wedge\bt^j_{ik}\wedge\bt^i_{jk} =
\left( (T_jQ_{ij}) (T_kQ_{jk}) (T_iQ_{ki}) - 
(T_jQ_{kj}) (T_kQ_{ik}) (T_iQ_{ji})
\right)  \bX_i \wedge  \bX_j \wedge  \bX_k, \qquad  i, j, k 
\quad \text{distinct}.
\end{equation*}
\end{proof}
\begin{Rem}
In the degenerate case, when 
a pair of opposite faces of the hexahedron is parallel, the corresponding
rotation coefficients vanish. Because they appear on different sides of equation 
\eqref{eq:CQL-constr}, then the constraint is automatically satisfied.
\end{Rem}
As it was shown in \cite{DS-sym}, the constraint \eqref{eq:CQL-constr} 
allows to rescale the forward and backward
data, using possibility given by equations
\eqref{eq:forward-scaling}-\eqref{eq:backward-scaling},  
to the form such that 
\begin{equation} \label{eq:symm-constr}
Q_{ij} = \tilde{Q}_{ij},  \qquad \text{or}
\qquad \rho_i T_i Q_{ji} = \rho_j T_j Q_{ij}, \qquad i\ne j,
\end{equation}
i.e., the rotation coefficients of CQL
are, in a sense, \emph{symmetric} with respect to
interchanging of their indices.
Under such a constraint equations \eqref{eq:rho-constr} and \eqref{eq:tau} 
allow to express the rotation coefficients in terms of the $\tau$-function 
as follows~\cite{Schief-JNMP}
\begin{equation}
(T_jQ_{ij})^2 = \frac{T_i \tau}{T_j \tau}
\left( 1 - \frac{(T_i T_j \tau) \tau}{(T_i \tau)( T_j \tau)} \right).
\end{equation}
Then the discrete
Darboux equations  equations \eqref{eq:MQL-Q} can be 
rewritten in the following quartic form
\begin{equation} \begin{split}
(\tau \,   T_i T_j T_k \tau -  T_i  \tau  \, T_j T_k \tau - 
T_j \tau  \, T_i T_k \tau -T_k \tau  \, T_i T_j & \tau )^2   = \\
4( T_i \tau  \, T_j \tau  \, T_i T_k \tau  \, T_j T_k \tau +  
T_i \tau  \, T_k \tau  \, T_i T_j \tau &  \, T_j T_k \tau + 
T_j \tau  \, T_k \tau  \, T_i T_k \tau  \, T_i T_j \tau - \\
& T_i \tau  \, T_j \tau  \, T_k \tau   \, T_i T_j T_k \tau - 
\tau  \, T_i T_j  \tau   \, T_j T_k \tau  \, T_i T_k \tau ), 
\end{split} \end{equation}
called in~\cite{Schief-JNMP} the discrete CKP equation.

Finally, we present a result which we will use in Section
\ref{sec:C-fund}. Its formal algebraic proof can be found in \cite{MM} but,
essentially, it uses the
facts that (i) the functions $\rho_i$ connect solutions of the forward and
backward linear problems, (ii) a solution of the adjoint linear problem 
\eqref{eq:lin-bH} is connected in this correspondence with solution of the
linear problem \eqref{eq:lin-X} but with backward rotation coefficients, and
(iii) backward and forward rotation coefficients in the CQL reduction coincide. 
\begin{Lem}[\cite{MM}] \label{lem-ry*-Y}
Given solution 
$\bY_i^*:\ZZ^N\to(\FF^K)^*$ of the adjoint linear problem \eqref{eq:lin-H}
for "symmetric" rotation coefficients \eqref{eq:symm-constr} then
$\rho_i (T_i\bY_i^*)^t:\ZZ^N\to\FF^K$ satisfies the corresponding linear 
problem \eqref{eq:lin-X}. 
\end{Lem}  

\subsection{Multidimensional consistency of the C-constraint}
\label{sec:M-cons-CQL}
As it was shown in \cite{MQL} the planarity condition, which allows to 
construct
the point $x_{123}$ as in Lemma \ref{lem:gen-hex}, does not lead to any further
restrictions if we increase dimension of the lattice. Because in the CQL case 
the constraint is imposed on the 3D (elementary hexahedra)
level, then to assure its multidimensional consistency we have to check the
four dimensional consistency. The multidimensional
consistency of the C-reduction with the geometric integrability scheme
would be the immediate consequence of its 4D consistency.

In fact, the consistency of the constraint has 
been proved algebraically
in \cite{DS-sym} starting from its algebraic form \eqref{eq:symm-constr}.
However, in that proof the main difficulty was shifted to the proof of existence
(in multidimensions) of the special choice of the $\tau$-function. In this paper
we present pure geometric proof of the 4D consistency of the CQL. We first
recall the relevant result on four dimensional consistency of the QL, which 
is the consequence of the of the following geometric observation.
\begin{Lem}[The 4D consistency of the geometric integrability scheme] 
\label{lem:4D-consist-QL}
Consider points $x_0$, $x_1$, $x_2$, $x_3$ and $x_4$
in general position in $\PP^M$, $M\geq 4$. Choose generic points 
$x_{ij}\in\langle x_0, x_i, x_j \rangle$, $1\leq i < j \leq 4$,
on the corresponding planes, and using
the planarity condition construct the points 
$x_{ijk}\in\langle x_0, x_i, x_j , x_k\rangle$, $1\leq i < j < k \leq 4$ -- the
remaining vertices of the four (combinatorial) cubes.
Then the intersection point $x_{1234}$ of the three planes 
\[\langle x_{12}, x_{123}, x_{124} \rangle, \;
\langle x_{13}, x_{123}, x_{134} \rangle, \;
\langle x_{14}, x_{124}, x_{134} \rangle \quad \text{in} \quad 
\langle x_{1}, x_{12}, x_{13}, x_{14} \rangle,
\] 
coincides with 
the intersection point of the three planes 
\[\langle x_{12}, x_{123}, x_{124} \rangle, \;
\langle x_{23}, x_{123}, x_{234} \rangle, \;
\langle x_{24}, x_{124}, x_{234} \rangle, \quad \text{in} \quad
\langle x_{2}, x_{12}, x_{23}, x_{24} \rangle,
\] 
which is the same as
the intersection point of the three planes 
\[\langle x_{13}, x_{123}, x_{134} \rangle, \;
\langle x_{23}, x_{123}, x_{234} \rangle, \;
\langle x_{34}, x_{134}, x_{234} \rangle, \quad \text{in} \quad 
\langle x_{3}, x_{13}, x_{23}, x_{34} \rangle,
\] 
and 
the intersection point of the three planes 
\[\langle x_{14}, x_{124}, x_{134} \rangle, \;
\langle x_{24}, x_{124}, x_{234} \rangle, \;
\langle x_{34}, x_{134}, x_{234} \rangle, \quad \text{in} \quad  
\langle x_{4}, x_{14}, x_{24}, x_{34} \rangle.
\]
\end{Lem}

\begin{Rem}
In fact, the point $x_{1234}$ is the unique
intersection point of the four three dimensional subspaces
$\langle x_{1}, x_{12}, x_{13}, x_{14} \rangle$,
$\langle x_{2}, x_{12}, x_{23}, x_{24} \rangle$,
$\langle x_{3}, x_{13}, x_{23}, x_{34} \rangle$,
and
$\langle x_{4}, x_{14}, x_{24}, x_{34} \rangle$ of the four dimensional subspace
$\langle x_{0}, x_{1}, x_{2}, x_{3} , x_{4} \rangle$. This observation
generalizes naturally to the case of more dimensional hypercube with the
planar facets.
\end{Rem}

It turns out
that the geometric core of the integrability of CQL is provided by
the Gallucci's theorem on eight skew lines (see, for example \cite{Coxeter-PG}).
\begin{Th}[The Gallucci theorem]
If three skew lines all meet three other skew lines, any transversal to the
first set of three meets any transversal to the second set.
\end{Th}
Motivated by the corresponding results
of \cite{KingSchief} we first prove a useful Lemma, 
where the Gallucci theorem is used.
\begin{Lem} \label{lem:C-consistency}
Consider four adjacent hexahedra with planar faces which share a vertex of a
4D hypercube. If three of them are C-hexahedra then 
the same holds also for the forth one.
\end{Lem}
\begin{proof}
We start from generic 4D hexahedron with planar faces. Denote by 
$\pi_{1234}$
the three dimensional subspace of $H_\infty$ being its intersection with the 4D
subspace $\langle x_0, x_1, x_2, x_3, x_4 \rangle$ of the hexahedron. By
$\pi_{ijk}$ denote the intersection line of the 3D subspace $\langle x_0, x_i,
x_j, x_k \rangle$ of one of the four 3D hexahedra with $H_\infty$. 

Denote by $h^k_{ij}$ the intersection points of the
opposite planes $\langle x_0, x_i, x_j \rangle$, 
$\langle x_k, x_{ik}, x_{jk} \rangle$ of the four
3D hexahedra with $H_\infty$
(all indices are distinct and range from $1$ to $4$, eventually
they should be reordered); these are the points entering into the definition of the
C-hexahedra. 
By $h^\ell_{ijk}$ denote the intersection lines of the 
three dimensional subspaces $\langle x_0, x_i, x_j, x_k \rangle$,  
$\langle x_\ell, x_{i\ell}, x_{j\ell}, x_{k\ell} \rangle$ and
$H_\infty$. There are four lines of this type (the $h$-family). 
Notice that point $h^k_{ij}$ is the intersection point of $h^k_{ij\ell}$ with
the plane $\pi_{ijk}$. On the plane $\pi_{ijk}$ we have therefore apart from
three collinear point $h^\ell_{ij}$, $h^\ell_{ik}$ and $h^\ell_{jk}$ (they
belong to $h^\ell_{ijk}$)
also three
other points $h^k_{ij}$, $h^j_{ik}$ and $h^i_{jk}$.

The C-reduction condition of the
hexahedron with with vertices $x_0$, $x_i$, $x_j$ and $x_k$ means collinearity 
of the points points $h^k_{ij}$, $h^j_{ik}$ and $h^i_{jk}$. If such a line 
exists we denote it by $g_{ijk}$.
It intersects not
only lines $h^k_{ij\ell}$, $h^j_{ik\ell}$ and $h^i_{jk\ell}$ (in points 
$h^k_{ij}$, $h^j_{ik}$ and $h^i_{jk}$, correspondingly), but also the fourth
line $h^\ell_{ijk}$ of the $h$-family, as both lines belong to the plane
$\pi_{ijk}$.

Let us assume that the C-reduction condition is satisfied for three hexahedra,
i.e., three such $g$-lines exist, all transversal to four $h$-lines. 
Assume that $g_{ijk}$, $g_{ij\ell}$ and $g_{ik\ell}$ exist,
define the line $G_{jk\ell}$ as the unique line passing through the points
$h^j_{k\ell}$ and $h^k_{j\ell}$, thus transversal to the lines
$h^j_{ik\ell}$ and $h^k_{ij\ell}$. 
Because $G_{jk\ell}$ is contained in $\pi_{jk\ell}$ then it must intersect 
also the line $h^i_{jk\ell}$ --- the third line of the $h$-family. 
By Gallucci's theorem it
intersects therefore the forth line $h^\ell_{ijk}$. 

That intersection point belongs to the plane $\pi_{jk\ell}$ (containing 
the points $h^j_{k\ell}$ and $h^k_{j\ell}$ which define $G_{jk\ell}$)
therefore it must be $h^\ell_{jk} = h^\ell_{ijk} \cap \pi_{jk\ell}$.
We have therefore shown that also the three points  
$h^j_{k\ell}$, $h^k_{j\ell}$ and $h^\ell_{jk}$ are collinear, i.e.,
$G_{jk\ell}= g_{jk\ell}$. 
\end{proof}
\begin{Prop}
Under hypotheses of Lemma~\ref{lem:gen-hex} assume that the C-reduction
condition is satisfied for 3D hexahedra meeting in one vertex of a 4D
hypercube.
Then the condition holds for all hexahedra of the hypercube.
\end{Prop}
\begin{proof}
Without loss of generality assume that the common vertex of the four hexahedra
(we know that it is enough to assume the C-reduction condition for three of
them) is $x_0$. Then three of them share the vertex $x_i$, which implies the
constraint for the forth one. All remaining
four hexahedra of the 4D hypercube which do
not share the point $x_0$ are of that type. 
\end{proof}

On the level of the 
multidimensional quadrilateral lattice there exists simple
alternative algebraic proof of an analogue of Lemma~\ref{lem:C-consistency},
which combined with ideas behind the proof of Proposition~\ref{prop:Q3-CQL}
would give the algebraic proof of the Lemma.
\begin{Cor}
Consider a 4D hexahedron withe vertices
$x$, $T_i x$, $T_j x$, $T_k x$,  $T_\ell x$ of the quadrilateral lattice 
$x:\ZZ^N\to\AAf^M$, $4\leq N \leq M$. If three of the four
3D hexahedra meeting in the vertex $x$ are the C-hexahedra then the 
same holds also for the forth one.
\end{Cor}
\begin{proof}
Assume that the fourth one is the hexahedron with basic vertices
$x$, $T_j x$, $T_k x$ and $T_\ell x$. 
By Proposition~\ref{prop:Q3-CQL} the C-reduction
condition for the remaining three reads
\begin{align}
(T_jQ_{ij}) (T_kQ_{jk}) (T_iQ_{ki}) &= (T_jQ_{kj}) (T_kQ_{ik}) (T_iQ_{ji}),\\
(T_i Q_{ji}) (T_j Q_{\ell j}) (T_\ell Q_{i\ell}) &= 
(T_jQ_{ij}) (T_\ell Q_{j\ell}) (T_iQ_{\ell i}),\\
(T_kQ_{ik}) (T_\ell Q_{k\ell}) (T_iQ_{\ell i}) &= 
(T_iQ_{ki}) (T_kQ_{\ell k}) (T_\ell Q_{i \ell}).
\end{align}
Multiplying the equations we obtain (assuming that the rotation coefficients do
not vanish) 
\begin{equation}
(T_kQ_{jk})(T_jQ_{\ell j})(T_\ell Q_{k\ell}) =
 (T_jQ_{kj})(T_\ell Q_{j\ell})(T_kQ_{\ell k}).
\end{equation}
\end{proof}

\section{Algebro-geometric construction of the C-quadrilateral lattice}
\label{sec:AG}
In this Section we apply the algebro-geometric approach, well known in the
theory of integrable systems \cite{BBEIM}, to the symmetric Darboux equations
and to the C-quadrilateral lattice.
Similar 
restrictions on the algebro-geometric data appeared in \cite{DJKM-CKP} in
construction of quasi periodic solutions of the CKP hierarchy. 
For discrete Laplace equations the algebro-geometric techniques were first
applied in \cite{Krichever-4p}. In \cite{AKV} the Egorov reduction of the
symmetric lattice \cite{Schief-Egorov} 
was studied by means of the algebro-geometric methods. In 1998 Peter Grinevich
isolated from \cite{AKV} these of the algebro-geometric conditions which
give rise the symmetric quadrilateral lattices \cite{PG-private}. 
Below we present that result on algebro-geometric
description of the C-(symmetric) quadrilateral lattice within broader context of
various discrete Baker--Akhiezer functions. 

In this Section we work over the field of complex numbers, i.e., $\FF=\CC$, and
we discuss the corresponding reality condintions. We mention that
many of the results gien below
can be transferred to the finite-field case, as in
\cite{BD,DBK}.

\subsection{The Baker--Akhiezer function of the quadrilateral lattice}
Let us consider a compact non-degenerate
Riemann surface $\mathcal{R}$ of genus $g$, a
non-special divisor $D=P_1 + \dots + P_g$ on $\mathcal{R}$, the
$N$ pairs of points $Q_i^\pm\in\mathcal{R}$, and the normalization
point $Q_\infty\in\mathcal{R}$. The system $\{ \mathcal{R}, D, Q_i^\pm, Q_\infty
\}$ is called \emph{the algebro-geometric data} used in the construction of the
quadrilateral lattice.
For simplicity we assume that all the
points above (including the points of $D$) are distinct. 

For $m\in\ZZ^N$ we define the meromorphic function 
$\psi(m):\mathcal{R}\to\CC\PP$ by prescribing
its analytical properties:\\
(i) as a function on $\mathcal{R}\setminus 
\cup_{i=1}^N \{ Q_i^\pm \}$ it may
may have as singularities only simple poles in points of the divisor 
$D$;\\
(ii) in points $Q_i^+$ (in points $Q_i^-$) it has poles (correspondingly, zeros)
of the order $m_i$, where by the pole of the negative
order we mean zero of the corresponding order;\\
(iii) in the point $Q_\infty$ the function $\psi$ is normalized to $1$.

By the standard (see \cite{AKV,BBEIM}) application of the Riemann--Roch theorem,
such a function exists and is unique.
When $z_i^\pm(P)$ is a local coordinate centered at $Q_i^\pm$ then 
$\psi$ in a neighbourhood of the point $Q_i^\pm$ is of the form
\begin{equation}
\psi(m|P) = \left(z_i^\pm(P)\right)^{\mp m_i}\left( \sum_{s=0}^\infty
\xi_{s}^{i,\pm}(m)\left(z_i^\pm(P)\right)^{s} \right).
\end{equation}  
By the standard reasoning in the finite-gap theory one can prove
the following result \cite{AKV} which connects the function $\psi$ with the
quadrilateral lattice theory.
\begin{Th}[\cite{AKV}]
For arbitrary point $P\in\mathcal{R}$ the Baker--Akhiezer
function $\psi$ satisfies 
in the variable
$m\in\ZZ^N$ the following system of linear equations
\begin{equation} \label{eq:Laplace-al-g}
\Delta_i\Delta_j\psi(m|P) = 
\left( T_i \frac{\Delta_j \xi_{0}^{i,+}(m)}{\xi_{0}^{i,+}(m)}\right)
\Delta_i\psi(m|P) +
\left( T_j \frac{\Delta_i \xi_{0}^{j,+}(m)}{\xi_{0}^{j,+}(m)} \right) 
\Delta_j\psi(m|P) ,\quad i\ne j.
\end{equation}
\end{Th}
\begin{Cor} \label{cor:measure}
Let $d {\boldsymbol{\mu}}(P)=(d\mu_1(P),\dots\, d\mu_M(P))^t$ 
be a vector valued measure on
$\mathcal{R}$, then the function $\bx:\ZZ^N\to\CC^M$ given by
\begin{equation} \label{eq:x-measure}
\bx(m) = \int_{\mathcal{R}} \psi(m|P) d {\boldsymbol{\mu}}(P),
\end{equation}
defines a quadrilateral lattice (in the complex affine space) 
with the functions $\xi_0^{i,+}$ as the Lam\'{e}
coefficients $H_i$.
\end{Cor}
\begin{Rem}
Similarly one shows hat $\bx$ satisfies the backward Laplace equations
\eqref{eq:Laplace-b} with the backward Lam\'e coefficients (proportional to) 
$\xi_0^{i,-}$; see also Proposition \ref{prop:t-psi}.
\end{Rem}
In order to have real quadrilateral 
lattices we must impose on the algebro-geometric data
certain reality restrictions. 
\begin{Cor} \label{cor:reality}
Assume that the Riemann surface $\mathcal{R}$ 
allows for an anti-holomorphic involution $\imath$. If
\begin{equation}
\imath(D) = D, \qquad \imath(Q_i^\pm) = Q_i^\pm, \qquad 
\imath(Q_\infty) = Q_\infty, \qquad d {\boldsymbol{\mu}}(\imath(P)) =
\overline{d {\boldsymbol{\mu}}P)},
\end{equation}
then the lattice $\bx$, given by equation \eqref{eq:x-measure} is real.
Moreover, if the local coordinate systems $z_i^\pm$ are compatible with
$\imath$, i.e.,
\begin{equation}
z_i^\pm(\imath(P)) = \overline{z_i^\pm(P)},
\end{equation}
then the Lam\'{e} coefficients, as defined above, are real functions.
\end{Cor}

\subsection{Other Baker--Akhiezer functions} 
In this Section we define
auxiliary Baker-Akhiezer functions which play the role of the forward
and backward normalized tangent vectors. Then we define the dual (adjoint)
Baker-Akhiezer function of the quadrilateral lattice.

\subsubsection{The forward Baker--Akhiezer functions $\psi_i$}
Below we consider new functions $\psi_i$
whose relation to $\psi$ is analogous to that between $\bX_i$ to $\bx$. 
Given $m\in\ZZ^N$ define \cite{AKV} the functions
$\psi_i(m)$ as meromorphic functions on $\mathcal{R}$ having the 
following analytic properties:\\
(i) as a function on $\mathcal{R}\setminus 
\cup_{i=1}^N \{Q_i^\pm\}$ it 
may have as singularities only simple poles in points of the divisor 
$D$;\\
(ii) in points $Q_j^+$ (in points $Q_j^-$) it has poles 
of the order $m_j+\delta_{ij}$ (correspondingly, zeros of the order $m_j$);\\
(iii) in the point $Q_\infty$ the function $\psi$ is equal to $0$.

By the Riemann--Roch theorem
the space of such functions is one dimensional. 
By choosing local coordinate $z_i^+$ near
$Q_i^+$ the function $\psi_i$ can be made unique by fixing its lowest order term at
$Q_i^+$ to one. Then near $Q_j^\pm$ we have the following local expansions
\begin{equation}
\psi_i(m|P) = \left(z_j^\pm(P)\right)^{\mp m_j}
\left( \frac{\delta_{ji}\delta_{\pm +}}{z_j^+(P)} + \sum_{s=0}^\infty
\zeta_{i,s}^{j,\pm}(m)\left(z_j^\pm(P)\right)^{s} \right).
\end{equation}
\begin{Th}[\cite{AKV}] \label{th:psi-i-AG}
The functions $\psi_i$ satisfy the equations
\begin{align} 
\label{eq:lin-x-X-AG}
\Delta_i\psi(m|P)&= (T_i\xi_{i,0}^{+}(m))\psi_i(m|P),\\
\label{eq:lin-X-AG}
\Delta_j\psi_i(m|P) &= (T_j\zeta_{i,0}^{j,+}(m))\psi_i(m|P),\qquad j\ne i,
\end{align}
whose expansion at $Q_k^+$, gives
\begin{align}
\Delta_i\xi_{k,0}^{+} & = (T_i\xi_{i,0}^{+}(m))\zeta_{i,0}^{k,+},
\qquad k\ne i,\\
\Delta_j\zeta_{i,0}^{k,+} & = (T_j\zeta_{i,0}^{j,+}) 
\zeta_{j,0}^{k,+} , \qquad k\ne j,
\end{align}
and allows for the identification
\begin{equation}
Q_{ij}(m) = \zeta_{i,0}^{j,+}(m).
\end{equation}
\end{Th}
\begin{Cor}
In notation of Corollary \ref{cor:measure} we have
\begin{equation} \label{eq:X-measure}
\bX_i(m) = \int_{\mathcal{R}} \psi_i(m|P) d {\boldsymbol{\mu}}(P).
\end{equation}
\end{Cor}
\begin{Rem}
Notice that different choices of local coordinates $z_i^+$ correspond to
rescaling of the forward data in agreement with equation \eqref{eq:forward-scaling}.
\end{Rem}
\subsubsection{The backward Baker--Akhiezer functions $\tilde\psi_i$}
We define the corresponding algebro-geometric
analog of the backward normalized tangent vectors
$\tbX_i$.
Given $m\in\ZZ^N$ define the functions
$\tilde\psi_i(m)$ as meromorphic functions on $\mathcal{R}$ having the 
following analytic properties:\\
(i) as a function on $\mathcal{R}\setminus 
\cup_{i=1}^N \{Q_i^\pm\}$ it may
may have as singularities only simple poles in points of the divisor 
$D$;\\
(ii) in points $Q_j^+$ (in points $Q_j^-$) it has poles 
of the order $m_j$ (correspondingly, zeros of the order $m_j-\delta_{ij}$);\\
(iii) in the point $Q_\infty$ the function $\psi$ is equal to $0$.

By the Riemann--Roch theorem the space of such functions is one dimensional. 
By choosing local coordinates $z_i^-$ near
$Q_i^-$ the function $\tilde\psi_i$ can be made unique by fixing its lowest order 
term at
$Q_i^-$ to one. Then near $Q_j^\pm$ we have the following local expansions
\begin{equation} \label{eq:exp-t-psi_i}
\tilde\psi_i(m|P) = \left(z_j^\pm(P)\right)^{\mp m_j}
\left( \frac{\delta_{ji}\delta_{\pm -}}{z_j^-(P)} + \sum_{s=0}^\infty
\tilde\zeta_{i,s}^{j,\pm}(m)\left(z_j^\pm(P)\right)^{s} \right).
\end{equation}
By the standard methods \cite{BBEIM} one can prove the following
analog of Theorem \ref{th:psi-i-AG}
\begin{Prop} \label{prop:t-psi}
The functions $\psi$ and $\tilde\psi_i$ are connected by the formulas
\begin{align}
\D_i\psi(m|P) & = -\left(T_i\tilde\psi_i(m|P)\right)\xi_{0}^{i,-}(m),
\label{eq:Ti-tpsi} \\
\D_j\tilde\psi_i(m|P) & = -\left(T_j\tilde\psi_j(m|P)\right) 
 \tilde\zeta_{i,0}^{j,-}(m) , \label{eq:Tj-tpsi-i}
\qquad j\ne i,
\end{align}
whose expansion at $Q_k^-$, gives
\begin{align}
\D_i\xi_{0}^{k,-} & = -\left(T_i\tilde\zeta_{i,0}^{k,-}\right)\xi_{0}^{i,-}, 
\qquad k\ne i,\\
\D_j \tilde\zeta_{i,0}^{k,-} &= - (T_j \tilde\zeta_{j,0}^{k,-})
\tilde\zeta_{i,0}^{j,-}, \qquad k\ne j,
\end{align}
and allows for the identification
\begin{equation}
\tH_i(m) = - \xi_{0}^{i,-}(m), \qquad \tQ_{ij}(m) = -\tilde\zeta_{j,0}^{i,-}(m).
\end{equation}
\end{Prop}
\begin{Cor}
In notation of Corollary \ref{cor:measure} we have
\begin{equation} \label{eq:tX-measure}
\tbX_i(m) = \int_{\mathcal{R}} \tilde\psi_i(m|P) d {\boldsymbol{\mu}}(P).
\end{equation}
Moreover, by comparing the analytical properties the functions $\psi_i$ and
$\tilde\psi_i$ we obtain
\begin{align}
\psi_i(m|P)  & = \left(T_i\tilde\psi_i(m|P)\right)\zeta_{i,0}^{i,-}(m) ,\\
T_i\tilde\psi_i(m|P) & = \left(T_i\tilde\zeta_{i,0}^{i,+}(m)\right) \psi_i(m|P), 
\end{align}
which allows for the identification 
\begin{equation}
\rho_i(m) = \zeta_{i,0}^{i,-}(m) = \frac{1}{T_i\tilde\zeta_{i,0}^{i,+}}.
\end{equation}
\end{Cor}
\begin{Rem}
Notice that different choices of local coordinates $z_i^-$ correspond to
rescaling of the backward data given by equation \eqref{eq:backward-scaling}.
\end{Rem}
\begin{Rem}
We are not concerned here about explicit theta-function formulas for the
Baker--Akhiezer functions and related potentials, see however \cite{AKV}. In
particular, the $\tau$-function of the quadrilateral lattice is, essentially
 \cite{AD-Chicago}, 
the Riemann theta function. 
\end{Rem}

\subsubsection{The dual Baker--Akhiezer functions}
In definition of the dual (adjoint) Baker--Akhiezer function of the
quadrilateral lattice we use the idea applied in \cite{DKJM} to construction of
the adjoint Baker--Akhiezer function of the KP hierarchy.

Denote by $\omega_\infty$ the meromorphic differential
with the only singularity being the second order pole at
$Q_\infty$, and whose holomorphic part is normalized
by vanishing of $\omega_\infty$ at
points of the divisor $D$  
\begin{equation}
\omega_\infty(P_i)=0, \quad i=1,\dots,g.
\end{equation}
\begin{Rem}
By choosing a coordinate
system $z_\infty(P)$ centered at $Q_\infty$
the differential $\omega_\infty$ can be made unique by
fixing its singular part in $z_\infty(P)$ as
\begin{equation}
\omega_\infty(P) = \left( \frac{1}{z_\infty(P)^2} + O(1)\right)dz_\infty(P),
\end{equation}
but we will not use that in the sequel.
\end{Rem}
Denote by $D^*$ the divisor of other $g$ zeros of $\omega_\infty$, and use it
to define the dual Baker--Akhiezer function $\psi^*$ 
exchanging also the role of the points $Q_i^+$ and $Q_i^-$:\\
(i) as a function on $\mathcal{R}\setminus 
\cup_{i=1}^N \{ Q_i^\pm\}$ it may
may have as singularities only simple poles in points of the divisor 
$D^*$;\\
(ii) in points $Q_i^+$ (in points $Q_i^-$) it has zeros (correspondingly, poles)
of the order $m_i$;\\
(iii) in the point $Q_\infty$ the function $\psi^*$ is normalized to $1$.

Using the Riemann--Roch theorem one can show that such function 
$\psi^*(m|P)$ exists and is unique.
In a neighbourhood of the point $Q_i^\pm$ it is of the form
\begin{equation}
\psi^*(m|P) = \left(z_i^\pm(P)\right)^{\pm m_i}\left( \sum_{s=0}^\infty
\xi_{s}^{*i,\pm}(m)\left(z_i^\pm(P)\right)^{s} \right).
\end{equation}  
Using the similar procedure like in the previous section it can be shown that the
dual function $\psi^*$ satisfies the Laplace equations with Lam\'{e} coefficients
$\xi_{0}^{*i,-}$, and it satisfies the backward Laplace equations with the backward
Lam\'{e} coefficients
$\xi_{0}^{*i,+}$.
\begin{Rem}
The meromorphic differential form
\begin{equation}
\omega = \psi \psi^* \omega_\infty 
\end{equation}
is singular only at $Q_\infty$ with the
singularity being the second order pole.
By the residue theorem the integral of $\omega$ around a closed contour around
$Q_\infty$ vanishes, which is the quadrilateral lattice counterpart of the 
celebrated bilinear identity \cite{DKJM} on the
algebro-geometric level.
\end{Rem}

In analogy to the Baker--Akhiezer functions $\psi_i$ and $\tilde\psi_i$
we may define the corresponding
dual Baker--Akhiezer functions. In the sequel we will need the analog of
$\tilde\psi_i$, which is defined as follows.  
Given $m\in\ZZ^N$ define the functions
$\psi_i^*(m)$ as meromorphic functions on $\mathcal{R}$ having the 
following analytic properties:\\
(i) as a function on $\mathcal{R}\setminus 
\cup_{i=1}^N Q_i^\pm$ it may
may have as singularities only simple poles in points of the divisor 
$D^*$;\\
(ii) in points $Q_j^+$ (in points $Q_j^-$) it has zeros 
of the order $m_j-\delta_{ij}$ (correspondingly, poles of the order $m_j$);\\
(iii) in the point $Q_\infty$ the function $\psi^*_i$ is equal to $0$.

By the Riemann--Roch the space of such functions is one dimensional. 
By choosing local coordinates $z_i^+$ near
$Q_i^+$ the function $\psi_i^*$ can be made unique by fixing its lowest order term at
$Q_i^+$ to one. Then near $Q_j^\pm$ we have the following local expansions
\begin{equation} \label{eq:exp-psi_i^*}
\psi_i^*(m|P) = \left(z_j^\pm(P)\right)^{\pm m_j}
\left( \frac{\delta_{ji}\delta_{\pm +}}{z_j^+(P)} + \sum_{s=0}^\infty
\zeta_{i,s}^{*j,\pm}(m)\left(z_j^\pm(P)\right)^{s} \right).
\end{equation}
As before one can study the relation between $\psi$ and $\psi_i^*$. 
However, we will be interested in the following connection formulas 
between the
$\zeta$-coefficients of both functions.
\begin{Prop} \label{prop:omega-z-z}
Denote by $a_i^\pm$ the first coefficients of the expansion of
$\omega_\infty$ near points $Q_i^\pm$
\begin{equation}
\omega_\infty(P) = [a_i^\pm + O(z_i^\pm(P))]dz_i^\pm(P),
\end{equation}
then by vanishing of the sum of residues of the differential
$\psi_i(m|P)\psi_j^*(m|P)\omega_\infty$ we have
\begin{equation} \label{eq:omega-z-z}
a_i^\pm\zeta_{j,0}^{*i,\pm}(m) + a_j^\pm\zeta_{i,0}^{j,\pm}(m) = 0, \qquad
i \neq j.
\end{equation}
\end{Prop}

\subsection{The algebro-geometric C-quadrilateral lattices}
Finally, we show that under certain restrictions on the algebro-geometric data
the finite-gap construction gives C-reduced quadrilateral lattice. This type of
restrictions appeared in \cite{DJKM-CKP} in construction of quasi-periodic
solutions of the CKP hierarchy. 
\begin{Prop} \label{prop:CQL-alg-geom}
Assume that $\mathcal{R}$ is equipped with the holomorphic involution
$\sigma:\mathcal{R}\to\mathcal{R}$ such that
\begin{equation}
\sigma(D^*) = D, \qquad \sigma(Q_i^\pm) = Q_i^\mp, \qquad 
\sigma(Q_\infty) = Q_\infty,
\end{equation}
then 
\begin{align}
\label{eq:psi-sigma}
\psi\circ\sigma & = \psi^*,\\
\label{eq:psi_i-sigma}
\tilde\psi_i\circ\sigma & = c_i \psi^*_i, \qquad c_i\in\CC.
\end{align}
\end{Prop}
\begin{proof}
In the standard way we compare analytic properties of both sides of each
equation. The
function $\psi\circ\sigma$ has the following analytic properties:\\
(i) as a function on $\mathcal{R}\setminus 
\cup_{i=1}^N \{Q_i^\pm\}$ it 
may have as singularities only simple poles in points of the divisor 
$D^*$;\\
(ii) in points $Q_i^+$ (in points $Q_i^-$) it has zeros (correspondingly, poles)
of the order $m_i$;\\
(iii) in the point $Q_\infty$ the function $\psi\circ\sigma$ is normalized to 
$1$.\\
Comparison with the analytic properties of $\psi^*$ and the Riemann--Roch
theorem gives equation \eqref{eq:psi-sigma}. 

Let us describe the analytic properties of the superposition
$\tilde\psi_i\circ\sigma$:\\
(i) as a function on $\mathcal{R}\setminus 
\cup_{i=1}^N \{Q_i^\pm\}$ it may
may have as singularities only simple poles in points of the divisor 
$D^*$;\\
(ii) in points $Q_j^+$ (in points $Q_j^-$) it has zeros 
of the order $m_j-\delta_{ij}$ (correspondingly, poles of the order $m_j$);\\
(iii) in the point $Q_\infty$ the function 
$\tilde\psi_i\circ\sigma$ is equal to $0$.\\
Therefore the function $\tilde\psi_i\circ\sigma$ must be proportional to
$\psi^*_i$.
\end{proof}
\begin{Cor}
Notice that 
when the local coordinates, which fix normalization of the functions, 
are chosen in agreement with the involution $\sigma$
\begin{equation} \label{eq:zi-zi+red}
z_i^-(\sigma(P)) = z_i^+(P),
\end{equation}
then the proportionality in equation
\eqref{eq:psi_i-sigma} becomes equality (i.e., $c_i=1$). Moreover, under such
conditions the
expansions \eqref{eq:exp-t-psi_i} and \eqref{eq:exp-psi_i^*} give
\begin{equation} \label{eq:tz-*z}
\tilde\zeta_{i,s}^{j,\pm}(m) = \zeta_{i,s}^{*j,\mp}(m).
\end{equation}
\end{Cor}
\begin{Th}[\cite{PG-private}]
Under assumptions of Proposition \ref{prop:CQL-alg-geom}, the quadrilateral
lattice constructed according to Corollary \ref{cor:measure} is subject to
the C-(symmetric) reduction.  
\end{Th}
\begin{proof}
Assume for a time being that the local coordinates $z_i^+$ are chosen in such  a
way that the first coefficients $a_i^+$ of the expansion of
$\omega_\infty$ near points $Q_i^+$ are equal (see Proposition
\ref{prop:omega-z-z}), and the local coordinates $z_i^-$ are chosen according to
equation \eqref{eq:zi-zi+red}; we may think of this special choice as using the
allowed freedom \eqref{eq:forward-scaling}-\eqref{eq:backward-scaling} in
definition of the backward and forward data.
Then equations \eqref{eq:omega-z-z} and
\eqref{eq:tz-*z} 
imply that
\begin{equation}
\zeta_{i,0}^{j,+}(m) = -\tilde\zeta_{j,0}^{i,-}(m) , \qquad \text{or} \quad
Q_{ij} = \tilde{Q}_{ij}.
\end{equation}
\end{proof}

\section{Transformation of the C-quadrilateral lattice}
\label{sec:C-fund}
We introduce geometrically the C-reduction
of the fundamental transformation of the quadrilateral lattice.  We also
connect this definition with earlier algebraic results of \cite{MM}. 
Then we prove the
corresponding permutability theorem for this transformation. 
\subsection{The vectorial fundamental
transformation of the quadrilateral lattice}
Let us first recall some basic facts concerning the vectorial fundamental
transformation of the quadrilateral lattice.
Geometrically, the (scalar) fundamental transformation is the relation 
between two 
quadrilateral
lattices $x$ and $x^\prime$ such that for each direction $i$ the
points $x$, $x^\prime$, $x_{(i)}$ and $x^\prime_{(i)}$ are coplanar. 

We
present below the algebraic description of its vectorial extension (see
\cite{MDS,TQL,MM} for details) in the affine formalism.
Given the solution $\boldsymbol{Y}_i:\mathbb{Z}^N\to\FF^K$, 
of the linear system \eqref{eq:lin-X}, and given the solution 
$\boldsymbol{Y}^*_i:\mathbb{Z}^N\to(\FF^K)^*$, of the linear system 
\eqref{eq:lin-H}. These allow to
construct the linear operator valued potential 
$\boldsymbol{\Omega}(\boldsymbol{Y},\boldsymbol{Y}^*):
\mathbb{Z}^N\to M^K_K(\FF)$,
defined by 
\begin{equation} \label{eq:Omega-Y-Y}
\Delta_i \boldsymbol{\Omega}(\boldsymbol{Y},\boldsymbol{Y}^*) = 
\boldsymbol{Y}_i \otimes T_i\boldsymbol{Y}^*_i, 
\qquad i = 1,\dots , N;
\end{equation} 
similarly, one defines 
$\boldsymbol{\Omega}(\boldsymbol{X},\boldsymbol{Y}^*):
\mathbb{Z}^N\to M^M_K(\FF)$ and 
$\boldsymbol{\Omega}(\boldsymbol{Y},H):
\mathbb{Z}^N\to \FF^K$ by
\begin{align} \label{eq:Omega-X-Y}
\Delta_i \boldsymbol{\Omega}(\boldsymbol{X},\boldsymbol{Y}^*) & = 
\boldsymbol{X}_i \otimes T_i \boldsymbol{Y}^*_i, \\
\Delta_i\boldsymbol{\Omega}(\boldsymbol{Y},H)  & =
\boldsymbol{Y}_i \otimes T_i H_i. 
\end{align} 
\begin{Prop}
If $\boldsymbol{\Omega}(\boldsymbol{Y},\boldsymbol{Y}^*)$ is invertible then
the vector function $\bx^\prime:\ZZ^N\to\FF^M$ given by
\begin{equation} \label{eq:fund-vect}
\bx^\prime  = \bx - 
\boldsymbol{\Omega}(\boldsymbol{X},\boldsymbol{Y}^*)
\boldsymbol{\Omega}(\boldsymbol{Y},\boldsymbol{Y}^*)^{-1}
\boldsymbol{\Omega}(\boldsymbol{Y},H),
\end{equation}
represents a quadrilateral lattice (the fundamental transform of $x$),
whose Lam\'e coefficients $H_i^\prime$, normalized tangent vectors
$\bX_i^\prime$ and rotation coefficients $Q_{ij}^\prime$ are given by
\begin{align} 
\label{eq:fund-vect-H}
H_i^\prime  &= H_i - 
\boldsymbol{Y}^*_i
\boldsymbol{\Omega}(\boldsymbol{Y},\boldsymbol{Y}^*)^{-1}
\boldsymbol{\Omega}(\boldsymbol{Y},H),\\
\label{eq:fund-vect-X}
\bX^\prime_i  & = \bX_i - 
\boldsymbol{\Omega}(\boldsymbol{X},\boldsymbol{Y}^*)
\boldsymbol{\Omega}(\boldsymbol{Y},\boldsymbol{Y}^*)^{-1}
\boldsymbol{Y}_i,\\
\label{eq:fund-vect-Q}
Q_{ij}^\prime  & = Q_{ij} - 
\boldsymbol{Y}^*_j
\boldsymbol{\Omega}(\boldsymbol{Y},\boldsymbol{Y}^*)^{-1}
\boldsymbol{Y}_i.
\end{align}
Moreover \cite{MM}, the  connection coefficients $\rho_i$ 
and the $\tau$-function
transform according
to
\begin{align} 
\label{eq:fund-vect-rho}
\rho^\prime_i & = \rho_i(1 + T_i \boldsymbol{Y}^*_i
\boldsymbol{\Omega}(\boldsymbol{Y},\boldsymbol{Y}^*)^{-1}
\boldsymbol{Y}_i),\\
\tau^\prime & = \tau \det\boldsymbol{\Omega}(\boldsymbol{Y},\boldsymbol{Y}^*).
\end{align}
\end{Prop}
The vectorial
fundamental transformation can be considered as superposition of
$K$ (scalar) fundamental transformations; on intermediate stages
the rest of the transformation data should be suitably transformed as well.
Such a description contains already the principle of permutability of such
transformations, which follows from the following observation~\cite{TQL}.
\begin{Prop}
Assume the following splitting of the data of the vectorial fundamental
transformation
\begin{equation}
\boldsymbol{Y}_i = \left( \begin{array}{c} 
\boldsymbol{Y}_i^a \\ \boldsymbol{Y}_i^b \end{array} \right),\qquad
\boldsymbol{Y}_i^* = \left( \begin{array}{cc} 
\boldsymbol{Y}_{ai}^{*}  & \boldsymbol{Y}_{b i}^{*} \end{array} \right),
\end{equation}
associated with the partition $\FF^K = \FF^{K_a} \oplus \FF^{K_b}$,
which implies the following splitting of the potentials
\begin{equation} \label{eq:split-fund-1}
\boldsymbol{\Omega}(\boldsymbol{Y},H) =  \left( \begin{array}{c} 
\boldsymbol{\Omega}(\boldsymbol{Y}^a,H) \\ 
\boldsymbol{\Omega}(\boldsymbol{Y}^b,H) \end{array} \right), \qquad
\boldsymbol{\Omega}(\boldsymbol{Y},\boldsymbol{Y}^*) = \left( \begin{array}{cc}
\boldsymbol{\Omega}(\boldsymbol{Y}^a,\boldsymbol{Y}_a^*) &
\boldsymbol{\Omega}(\boldsymbol{Y}^a,\boldsymbol{Y}_b^*) \\
\boldsymbol{\Omega}(\boldsymbol{Y}^b,\boldsymbol{Y}_a^*) &
\boldsymbol{\Omega}(\boldsymbol{Y}^b,\boldsymbol{Y}_b^*)\end{array} \right), 
\end{equation}
\begin{equation} \label{eq:split-fund-2}
\boldsymbol{\Omega}(\boldsymbol{X},\boldsymbol{Y}^*) = \left( \begin{array}{cc} 
\boldsymbol{\Omega}(\boldsymbol{X},\boldsymbol{Y}_a^*)  &
\boldsymbol{\Omega}(\boldsymbol{X},\boldsymbol{Y}_b^*)\end{array} \right).
\end{equation} 
Then the vectorial fundamental transformation is equivalent to the following
superposition of vectorial fundamental transformations:\\
1) Transformation $\bx\to\bx^{\{a\}}$ with the data 
$\boldsymbol{Y}_i^a$, $\boldsymbol{Y}_{ai}^*$ and the corresponding
potentials
$\boldsymbol{\Omega}(\boldsymbol{Y}^a,H)$, 
$\boldsymbol{\Omega}(\boldsymbol{Y}^a,\boldsymbol{Y}_a^*)$, 
$\boldsymbol{\Omega}(\boldsymbol{X},\boldsymbol{Y}_a^*)$
\begin{align}
\label{eq:fund-vect-a}
\bx^{\{a\}}  & = \bx - 
\boldsymbol{\Omega}(\boldsymbol{X},\boldsymbol{Y}^*_a)
\boldsymbol{\Omega}(\boldsymbol{Y}^a,\boldsymbol{Y}^*_a)^{-1}
\boldsymbol{\Omega}(\boldsymbol{Y}^a,H),\\
\boldsymbol{X}_i^{\{a\}} & = \boldsymbol{X}_i -
\boldsymbol{\Omega}(\boldsymbol{X},\boldsymbol{Y}^*_a)
\boldsymbol{\Omega}(\boldsymbol{Y}^a,\boldsymbol{Y}^*_a)^{-1}
\boldsymbol{Y}^a_i,
\\
H_i^{\{a\}} & = H_i - \boldsymbol{Y}^*_{i a}
\boldsymbol{\Omega}(\boldsymbol{Y}^a,\boldsymbol{Y}^*_a)^{-1}
\boldsymbol{\Omega}(\boldsymbol{Y}^a,H).
\end{align}
2) Application on the result the vectorial fundamental transformation with the
transformed data
\begin{align}
{\boldsymbol{Y}}_i^{b\{a\}} & = \boldsymbol{Y}_i^b -
\boldsymbol{\Omega}(\boldsymbol{Y}^b,\boldsymbol{Y}^*_a)
\boldsymbol{\Omega}(\boldsymbol{Y}^a,\boldsymbol{Y}^*_a)^{-1}
\boldsymbol{Y}^a_i,
\\
{\boldsymbol{Y}}_{i b}^{*\{a\}} & = \boldsymbol{Y}_{i b}^* - 
\boldsymbol{Y}^*_{i a}
\boldsymbol{\Omega}(\boldsymbol{Y}^a,\boldsymbol{Y}^*_a)^{-1}
\boldsymbol{\Omega}(\boldsymbol{Y}^a, \boldsymbol{Y}_{b}^*),
\end{align}
and potentials
\begin{align} 
{\boldsymbol{\Omega}}(\boldsymbol{Y}^b,H)^{\{a\}} & =
\boldsymbol{\Omega}(\boldsymbol{Y}^b,H) - 
\boldsymbol{\Omega}(\boldsymbol{Y}^b,\boldsymbol{Y}^*_a)
\boldsymbol{\Omega}(\boldsymbol{Y}^a,\boldsymbol{Y}^*_a)^{-1}
\boldsymbol{\Omega}(\boldsymbol{Y}^a,H)=
\boldsymbol{\Omega}({\boldsymbol{Y}}^{b\{a\}},H^{\{a\}}),
\\
{\boldsymbol{\Omega}}(\boldsymbol{Y}^b,\boldsymbol{Y}^*_b)^{\{a\}} & =
\boldsymbol{\Omega}(\boldsymbol{Y}^b,\boldsymbol{Y}^*_b) - 
\boldsymbol{\Omega}(\boldsymbol{Y}^b,\boldsymbol{Y}^*_a)
\boldsymbol{\Omega}(\boldsymbol{Y}^a,\boldsymbol{Y}^*_a)^{-1}
\boldsymbol{\Omega}(\boldsymbol{Y}^a,\boldsymbol{Y}^*_b)=
\boldsymbol{\Omega}({\boldsymbol{Y}}^{b\{a\}},{\boldsymbol{Y}}_b^{*\{a\}}),
\\
{\boldsymbol{\Omega}}(\boldsymbol{X},\boldsymbol{Y}^*_b)^{\{a\}}  & =
\boldsymbol{\Omega}(\boldsymbol{X},\boldsymbol{Y}^*_b) - 
\boldsymbol{\Omega}(\boldsymbol{X},\boldsymbol{Y}^*_a)
\boldsymbol{\Omega}(\boldsymbol{Y}^a,\boldsymbol{Y}^*_a)^{-1}
\boldsymbol{\Omega}(\boldsymbol{Y}^a,\boldsymbol{Y}^*_b)=
\boldsymbol{\Omega}({\boldsymbol{X}}^{\{a\}},{\boldsymbol{Y}}_b^{*\{a\}}),
\label{eq:fund-vect-potentials-slit}
\end{align}
i.e.,
\begin{equation} \label{eq:fund-vect-a-b}
\bx^\prime = \bx^{\{a,b\}}  = \bx^{\{a\}} - 
{\boldsymbol{\Omega}}(\boldsymbol{X},\boldsymbol{Y}^*_b)^{\{a\}}
[{\boldsymbol{\Omega}}(\boldsymbol{Y}^b,\boldsymbol{Y}^*_b)^{\{a\}}]^{-1}
{\boldsymbol{\Omega}}(\boldsymbol{Y}^b,H)^{\{a\}}.
\end{equation}
\end{Prop}
\begin{Rem}
The same result $\bx^\prime = \bx^{\{a,b\}}={\bx}^{\{b,a\}}$
is obtained exchanging the order of transformations, exchanging also the indices
$a$ and $b$ in formulas 
\eqref{eq:fund-vect-a}-\eqref{eq:fund-vect-a-b}. 
\end{Rem}

\subsection{The CQL (symmetric) reduction of the fundamental transformation}
In this section we describe restrictions on the data of the fundamental
transformation in order to preserve the reduction from QL to CQL. 
As usually (see, for example \cite{TQL,q-red,BQL}) a reduction of the
fundamental transformation for a special quadrilateral lattice 
mimics the geometric properties of the lattice. Because the basic geometric
property of the (scalar) fundamental transformation can be interpreted as
construction of a "new level" of the quadrilateral lattice, then it is natural 
to define the reduced transformation in a similar spirit. Our definition of  
\emph{the CQL reduction of the fundamental transformation} is therefore based on
the following observation. 
\begin{Lem} \label{lem:BQL-fund}
Given quadrilateral lattice $x:\ZZ^N\to\AAf^M$ and its fundamental transform
$x^\prime$ constructed under additional assumption that for any point
$x$ of the lattice and any pair $i,j$ of different directions, the hexahedra
with basic vertices
$x$, $T_i x$, $T_j x$ and $x^\prime$ satisfy the C-reduction condition. 
Then both the starting lattice $x:\ZZ^N\to\AAf^M$ and its
transform $x^\prime:\ZZ^N\to\AAf^M$ are C-quadrilateral lattices.
\end{Lem}
\begin{proof}
As $N\geq 3$, by 
Lemma~\ref{lem:C-consistency} we have that also the hexahedra
with basic vertices
$x$, $T_i x$, $T_j x$ and $T_k x$, with $i,j,k$ distinct, 
satisfy the C-reduction condition. The similar statement for the transformed
lattice is a consequence of 
the $4$-dimensional consistency of the CQL lattice.
\end{proof}
\begin{Def}
The fundamental transform $x^\prime$ of a C-quadrilateral lattice 
$x:\ZZ^N\to\AAf^M$  
constructed under additional assumption that for any point $x$ of the lattice 
and any pair $i,j$ of different directions, the hexahedra
with basic vertices
$x$, $T_i x$, $T_j x$ and $x^\prime$ satisfy the C-reduction condition, 
is called
\emph{the CQL reduction} of the fundamental transformation.
\end{Def}
The following result gives the corresponding
restriction of the data of the (scalar) fundamental transformation. 
\begin{Prop} \label{prop:CQL-fund}
Let $x$ be a C-quadrilateral lattice with rotation coefficients
satisfying constraint \eqref{eq:symm-constr}, and $x^\prime$ its 
C-reduced fundamental transform. Then there exists a constant $c$ such that 
the data 
$Y_i:\ZZ^N\to \FF$ and $Y_i^*:\ZZ^N\to\FF$ of the transformation are connected
by relation
\begin{equation}
Y_i = c \, \rho_i T_i Y_i^*, \qquad i = 1,\dots , N, \qquad c\in\FF.
\end{equation}
\end{Prop}
\begin{proof}
We start from considerations similar to that of proof of 
Proposition~\ref{prop:Q3-CQL}. The idea is to interpret 
the fundamental transformation as construction of a new
level of the quadrilateral lattice.  The potential
$\boldsymbol{\Omega}(\boldsymbol{X},Y^*)$, called also the Combescure vector of
the transformation, serves as the normalized tangent vector \cite{TQL}, 
which we denote by
$\bX_\prime$, in the transformation direction ``$\prime$".

Denote by $\bt^\prime_{ij}$ ($i,j$ are
distinct) the direction vector of the common line of the planes 
$\langle \bx, T_i \bx, T_j\bx \rangle$ and  
$\langle \bx^\prime, T_i \bx^\prime, T_j \bx^\prime \rangle$. 
It must be therefore decomposed in
the basis $\{ \bX_i,\bX_j \}$ and in the basis 
$\{ \bX_i^\prime,\bX_j^\prime \}$.
Assuming its decomposition in the second basis we get
\begin{equation*}
\bt^\prime_{ij} = a \bX^\prime_i + b \bX_j^\prime = a \bX_i + b \bX_j -
\bX_\prime(a Y_i + b Y_j ) \frac{1}{\boldsymbol{\Omega}(Y,Y^*)},
\end{equation*}
where we have used the transformation equation \eqref{eq:fund-vect-X}. 
Because the coefficient
in front of $\bX_\prime$ must vanish, the vector can be therefore
chosen as
\begin{equation}
\bt^\prime_{ij}  = Y_j \bX_i - Y_i \bX_j . 
\end{equation}

Similarly, denote by $\bt^j_{\prime i}$ 
the direction vector of the intersection line of the plane 
$\langle \bx, \bx^\prime, T_i\bx \rangle$ with  
$\langle T_j \bx, T_j \bx^\prime, T_i T_j \bx\rangle$. 
It must be therefore decomposed in
the basis $\{ \bX_\prime, \bX_i \}$ and in the basis 
$\{ T_j \bX_\prime, T_j\bX_i \}$.
By using equations \eqref{eq:Omega-X-Y} and \eqref{eq:lin-X} we can 
choose the vector as
\begin{equation}
\bt^\prime_{ij}  = (T_j Q_{ij}) T_j \bX_\prime - (T_j Y_j^*) T_j\bX_i =
(T_j Q_{ij}) \bX_\prime - (T_j Y_j^*)  \bX_i. 
\end{equation}
Because
\begin{equation*}
\bt^j_{\prime i}\wedge\bt^i_{\prime j}\wedge\bt^\prime_{ij} =
\left( (T_jQ_{ij}) Y_j (T_iY^*_i) - 
(T_iQ_{ji}) Y_i (T_j Y^*_j)
\right)  \bX_\prime \wedge \bX_i \wedge  \bX_j \qquad  i\ne j,
\end{equation*}
then the C-reduction condition of the $i$, $j$, $\prime$ hexahedron takes the
form of equation \eqref{eq:CQL-constr}
\begin{equation}
(T_i Q_{ji})  Y_j (T_iY^*_i)=
(T_j Q_{ij})Y_i (T_j Y^*_j). 
\end{equation}
Making use of condition \eqref{eq:symm-constr}  we obtain that
\begin{equation}
\rho_j (T_j Y^*_j) Y_i = \rho_i (T_i Y^*_i) Y_j .
\end{equation}
Then we use Lemma \ref{lem-ry*-Y}, which states that 
$\rho_i T_i Y_i^*$ satisfies the same linear problem
\eqref{eq:lin-X} as $Y_i$ does.  
Finally, application of 
the following Lemma concludes the proof. 
\end{proof}
\begin{Lem}
Any two scalar solutions $Y_i$ and $\tilde{Y}_i$ of the linear problem 
\eqref{eq:lin-X}, which satisfy the constraint
\begin{equation} \label{eq:hY-Y}
\hat{Y}_j Y_i = \hat{Y}_i Y_j,
\end{equation}
must be proportional.
\end{Lem}
\begin{proof}[Proof of the Lemma]
Assume that none of the solutions is trivial (then the proportionality constant
would be zero) and define
\begin{equation}
r_i = \frac{\hat{Y}_i}{Y_i}.
\end{equation}
By equation \eqref{eq:lin-X} we find
\begin{equation}
\Delta_j r_i = \frac{(T_j Q_{ij})(\hat{Y}_j Y_i - \hat{Y}_i Y_j)}
{Y_i (T_j Y_i)}, \qquad j \ne i,
\end{equation}
which vanishes because of the assumption \eqref{eq:hY-Y}, therefore $r_i$ may
depend on the variable $m_i$ only. Inserting then $\hat{Y}_i = r_i Y_i$ into
equation \eqref{eq:hY-Y} we obtain $r_i = r_j$, which implies that all the $r$'s
are equal to the same constant. 
\end{proof}
\begin{Rem}
In the non-degenerate situation, i.e. $c\ne 0$, which we assume in the sequel, 
we can put $c=1$, because (up to initial value)
$\boldsymbol{\Omega}(cY,{Y}^*) = c\boldsymbol{\Omega}(Y,{Y}^*)$,
and 
$\boldsymbol{\Omega}(cY,H)= c\boldsymbol{\Omega}(Y,H)$, and the final result 
\eqref{eq:fund-vect} is independent of $c$.
\end{Rem}
\subsection{Permutability theorem for 
the CQL reduction of the fundamental
transformation}
In this section we study restrictions of the data of the vectorial fundamental
transformation, which are compatible with the CQL reduction. In that part we
follow the corresponding results of \cite{MM} (see also Proposition 4.9 of
\cite{DS-sym}). 
Then we show the
corresponding permutability property of the transformation. 

\begin{Prop}[\cite{MM}]
Given a solution $\bY_i^*$ of the adjoint linear problem \eqref{eq:lin-H} for
the C-quad\-ri\-la\-te\-ral lattice whose rotation coefficients satisfy
the CQL constraint \eqref{eq:symm-constr} then 
\begin{equation} \label{eq:C-ft-YY}
\bY_i = \rho_i (T_i \bY_i^*)^t
\end{equation}
provides a vectorial solution of the linear problem 
\eqref{eq:lin-X}, and the corresponding potential
$\boldsymbol{\Omega}(\boldsymbol{Y},\boldsymbol{Y}^*) $ allows for the following
constraint
\begin{equation}\label{eq:C-ft-O}
\boldsymbol{\Omega}(\boldsymbol{Y},\boldsymbol{Y}^*)^t=
\boldsymbol{\Omega}(\boldsymbol{Y},\boldsymbol{Y}^*).
 \end{equation}
 With such a data the transformed lattice $\bx^\prime$ given by
 \eqref{eq:fund-vect} is C-quadrilateral lattice as well.
\end{Prop}

\begin{Rem}
In \cite{MM}, instead of relation \eqref{eq:C-ft-YY} it was used more general
relation
\begin{equation} 
\hat\bY_i = A \rho_i (T_i \bY_i^*)^t,
\end{equation}
where $A$ is an arbitrary linear operator. Then also the constraint 
\eqref{eq:C-ft-O} had to be replaced by
\begin{equation} \label{eq:C-ft-AO}
A\boldsymbol{\Omega}(\hat{\boldsymbol{Y}},\boldsymbol{Y}^*)^t=
\boldsymbol{\Omega}(\hat{\boldsymbol{Y}},\boldsymbol{Y}^*) A^t,
\end{equation}
which is however equivalent, up to initial data, 
to \eqref{eq:C-ft-O} due to
\begin{equation}
A\boldsymbol{\Omega}(A\boldsymbol{Y},\boldsymbol{Y}^*)^t -
\boldsymbol{\Omega}(A\boldsymbol{Y},\boldsymbol{Y}^*) A^t =
A\left(  \boldsymbol{\Omega}(\boldsymbol{Y},\boldsymbol{Y}^*)^t -
\boldsymbol{\Omega}(\boldsymbol{Y},\boldsymbol{Y}^*) \right) A^t .
\end{equation}
Moreover, because (up to initial value)
$\boldsymbol{\Omega}(A\bY,{\bY}^*) = A\boldsymbol{\Omega}(\bY,{\bY}^*)$,
and 
$\boldsymbol{\Omega}(A\bY,H)= A\boldsymbol{\Omega}(\bY,H)$ the final result 
\eqref{eq:fund-vect} is independent of (non-degenerate) $A$. 
\end{Rem}

\begin{Prop}
The fundamental vectorial transform given by
\eqref{eq:fund-vect} with the data restricted by conditions
\eqref{eq:C-ft-YY} and \eqref{eq:C-ft-O}
can be considered as the superposition of $K$ (scalar)
discrete CQL reduced fundamental transforms. 
\end{Prop}
\begin{proof}
For $K=1$ we obtain the 
CQL reduction of the fundamental transformation in the
setting of Proposition~\ref{prop:CQL-fund} (with $c=1$). For $K>1$
the statement follows from the standard reasoning applied to superposition of
two reduced vectorial fundamental transformations (compare with
\cite{TQL,q-red,BQL}). 

Assume the splitting 
$\FF^K=\FF^{K_a}\oplus\FF^{K_b}$ 
and the induced splitting 
\begin{equation}
\boldsymbol{Y}_i^* = \left( \begin{array}{cc} 
\boldsymbol{Y}_{ai}^{*}\; & \boldsymbol{Y}_{b i}^{*} \end{array} \right),
\end{equation}
of the basic data $\boldsymbol{Y}_i^*$ of the
transformation. Then we have also 
\begin{equation}
\boldsymbol{Y}_i = \left( \begin{array}{c} 
\boldsymbol{Y}_i^a \\ \boldsymbol{Y}_i^b \end{array} \right) =
\left( \begin{array}{c} 
\rho_i(T_i\boldsymbol{Y}^*_{ai})^t \\ \rho_i(T_i\boldsymbol{Y}^*_{bi})^t 
\end{array} \right),
\end{equation}
and (in the shorthand notation, compare equations
\eqref{eq:split-fund-1}-\eqref{eq:split-fund-2})
\begin{equation}
\boldsymbol{\Omega}(\boldsymbol{Y},\boldsymbol{Y}^*) = 
\left(   \begin{array}{cc}
\boldsymbol{\Omega}^a_a &  \boldsymbol{\Omega}^a_b \\
\boldsymbol{\Omega}^b_a &  \boldsymbol{\Omega}^b_b 
\end{array} \right),
\end{equation}
while the constraint \eqref{eq:C-ft-O} reads
\begin{equation} \label{eq:C-ft-Os-split} 
(\boldsymbol{\Omega}^a_a)^t = \boldsymbol{\Omega}^a_a, \qquad
(\boldsymbol{\Omega}^a_b)^t =  \boldsymbol{\Omega}^b_a, \qquad  
(\boldsymbol{\Omega}^b_b)^t = \boldsymbol{\Omega}^b_b .
\end{equation}
By straightforward algebra, using equations 
\eqref{eq:C-ft-Os-split}, one checks that the transformed data
satisfy the CQL constraints \eqref{eq:C-ft-YY} and \eqref{eq:C-ft-O}
as well, i.e.,
\begin{equation}
\boldsymbol{Y}_i^{b \{ a\}} = \rho_i^{\{ a\}} 
(T_i\boldsymbol{Y}^{*\{ a\}}_{b i})^t ,\qquad
(\boldsymbol{\Omega}^{b \{ a\}}_b)^t = \boldsymbol{\Omega}^{b \{ a\}}_b,
\end{equation}
which concludes the proof.
\end{proof}
\begin{Rem}
In the case with matrix $A$ as in previous Remark, the scalar components of the
vectorial transformation do not satisfy (unless $A$ is diagonal) the CQL
reduction condition of Proposition~\ref{prop:CQL-fund}.
\end{Rem}
\begin{Rem}
Because the CQL-reduced fundamental transformation can be considered as
construction of new levels of the C-quadrilateral lattice, then
if we denote by $x^{\{1,2\}}$ the
C-quadrilateral lattice obtained by superposition of two (scalar) such
transforms from $x$ to $x^{\{1\}}$ and $x^{\{2\}}$, then 
for each direction $i$ of the lattice the hexahedra with basic vertices
$x$, $T_i x$, $x^{\{1\}}$ and $x^{\{2\}}$ are C-hexahedra.
Similarly, if we consider superpositions of three (scalar) transforms of the
C-quadrilateral lattice $x$ then  
the hexahedra with basic vertices
$x$, $x^{\{1\}}$, $x^{\{2\}}$ and $x^{\{3\}}$ are
C-hexahedra. 
\end{Rem}
\section{Conclusion and remarks}
We presented new geometric interpretation of the discrete CKP equation within
the theory of quadrilateral lattices. The paper should be considered as
supplementary to \cite{DS-sym}. It has been also written in a format similar
to \cite{BQL}, where we presented novel geometric interpretation of the discrete
BKP equation. 
Results of the paper show once again the
fundamental role of the incidence geometry structures
in the integrable geometry. We remark that the integrability
of the discrete BKP equations was a consequence of the M\"{o}bius theorem on
mutually inscribed tetrahedrons, and the integrability
of the discrete CKP equations was a consequence of the Gallucci theorem.
However, it turns out   \cite{Coxeter-PG} that
both theorems are two diferent faces of a more fundamental
result concerning the so called quadrangular
sets of points.

In the Appendix we present the theory of the Darboux maps
within the quadrilateral lattice theory thus showing the fundamental role of the
quadrilateral lattice in integrable discrete geometry.

\appendix
\section{The Darboux maps within the quadrilateral lattice theory}
We would like to present an interpretation of the so called Darboux maps
\cite{Schief-JNMP,KingSchief} within the quadrilateral lattice theory.  
Denote by $\EE =
\EE(\ZZ^N)$ the set of edges of the $\ZZ^N$ lattice. Consider a map
\begin{equation}
\boldsymbol{v}:\EE\to\RR^M,
\end{equation} 
regarded as a set of $N$ maps $\boldsymbol{v}^i:\ZZ^N\to\RR^M$ of edges in 
$i$th
direction. It is termed a discrete Darboux map if the four images of the edges
of any face of the $\ZZ^N$ lattice are collinear, i.e., there exist functions
$\rho_{ij}$, $i\ne j$, such that
\begin{equation} \label{eq:Darboux-map}
\D_j\boldsymbol{v}^i = (T_i\rho^{ij})
(T_i\boldsymbol{v}^j - T_j\boldsymbol{v}^i),
\qquad i\ne j. 
\end{equation}
Compatibility of equations \eqref{eq:Darboux-map} implies \cite{Schief-JNMP} 
that the functions $\rho^{ij}$ satisfy the discrete Darboux equations
\eqref{eq:MQL-A}.
\begin{Rem}
In order to use the Darboux equations in the form \eqref{eq:MQL-A} the
definition of $\rho^{ij}$ in this paper is shifted with respect to that used in
\cite{Schief-JNMP,KingSchief}.
\end{Rem}

\begin{figure}
\begin{center}
\includegraphics{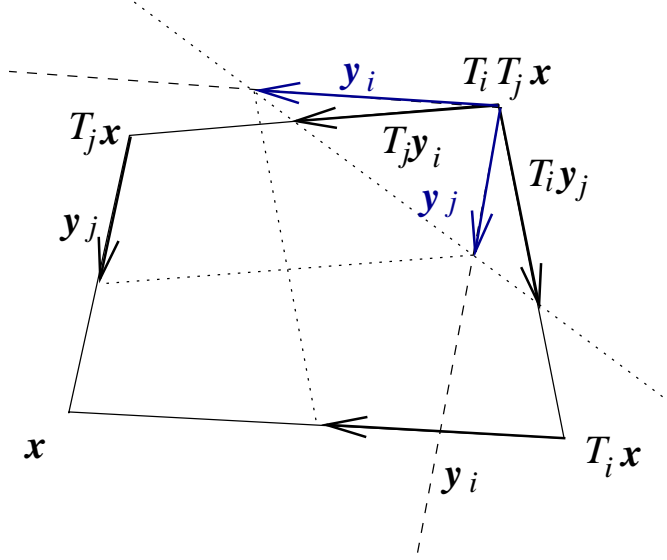}
\end{center}
\caption{The Darboux maps and the quadrilateral lattices}
\label{fig:Darboux-map}
\end{figure}

We will briefly demonstrate that the Darboux maps can be interpreted as suitably rescaled
normalized backward tangent vectors $\tbX_i$; 
compare Figure 
\ref{fig:back} with Figure \ref{fig:Darboux-map},
where also the geometric construction of the Darboux map is given.

\begin{Prop}
Consider the quadrilateral lattice $\bx:\ZZ^N\rightarrow\RR^M$ together with
its backward tangent vectors $\tbX_i$ and the corresponding backward rotation
coefficients $\tQ_{ij}$.
Let $\tv_i$ be a scalar solution of the backward linear problem 
\eqref{eq:lin-bX}
\begin{equation} \label{eq:lin-tv}
\tD_i\tv_j = (T_i^{-1} \tQ_{ij})\tv_i \; , \qquad \text{or}
\quad \D_i\tv_j =  (T_i\tv_i)\tQ_{ij},    \quad i\ne j \; ,
\end{equation}
define the maps $\by_i:\ZZ^N\to\RR^M$, $i=1,\dots , N$,
\begin{equation} \label{eq:y-tX}
\by_i = T_i\left( \frac{1}{\tv_i}\tbX_i \right).
\end{equation}
Then the maps $\by_i$ satisfy the Darboux map equations
\begin{equation} \label{eq:Darboux-map-vby}
\D_i \by_j = (T_j B_{ji})(T_j \by_i - T_i \by_j),
\end{equation}
with the coefficients
\begin{equation} \label{eq:Darboux-map-Lame}
B_{ij} = \frac{\D_j \tv_i}{\tv_i}, \qquad i\ne j, \quad i,j=1,\dots , N.
\end{equation}
\end{Prop}
\begin{proof}
By direct verification using the fact that
both $\tbX_i$ and $\tv_i$ satisfy the same linear system \eqref{eq:lin-bX}.
Geometrically, by results of Appendix A1 of \cite{DNS-Bianchi}, it means
that $\by_i$ and
$\by_j$ represent mutual Laplace transforms \cite{DCN} in the affine gauge, i.e.  
$\by_i$, $T_j\by_i$, $\by_j$ and $T_i\by_j$ are collinear and satisfy equation
of the form of \eqref{eq:Darboux-map-vby}.
\end{proof}
\begin{Cor}
The above result can be reversed, i.e., any Darboux map gives rise via equations 
\eqref{eq:y-tX} and \eqref{eq:Darboux-map-Lame} to a system of normalized
backward tangent vectors of a quadrilateral lattice. Thus the correspondence
between Darboux maps and quadrilateral lattices occurs on the geometric linear
level.
\end{Cor}
\begin{Rem}
Notice that the functions
$\tv_i$ satisfy the forward adjoint linear problem \eqref{eq:lin-H} with the
rotation coefficients $\tQ_{ij}$ which satisfy the MQL equations 
\eqref{eq:MQL-Q}. Then without any calculation we infer
that the coefficients $B_{ij}$
are solutions of the discrete Darboux (MQL) equations \eqref{eq:MQL-A}.
\end{Rem}
Finally, we mention that to the vectors $\by_i$ it can be given 
geometric meaning as non-homogeneous
coordinates in $H_\infty$ of the intersections points
$\langle x, T_i x \rangle \cap H_\infty$ of
the tangent lines to the quadrilateral lattice $x:\ZZ^N\to\PP^M$ with the
hyperplane at infinity $H_\infty$ (see Section \ref{sec:M-cons-CQL}).
Using the Pascal hexagon theorem it can be shown that within this interpretation
the "conic condition" of \cite{KingSchief} is equivalent to Definition
\ref{def:C-hexahedron} of the C-hexahedron. 

\section*{Acknowledgements}
The paper was supported by the Polish Ministry of  
Science and Higher Education research grant 1~P03B~017~28.

\bibliographystyle{amsplain}

\providecommand{\bysame}{\leavevmode\hbox to3em{\hrulefill}\thinspace}

\end{document}